\documentclass[
]{ceurart}

\usepackage{multirow}
\usepackage{subcaption}
\usepackage{times}
\usepackage{latexsym}
\usepackage{todonotes}
\usepackage{booktabs}
\usepackage{amssymb}
\usepackage{colortbl}
\usepackage{comment}
\usepackage[utf8]{inputenc}
\usepackage[T1]{fontenc}

\usepackage{microtype}

\usepackage{inconsolata}

\usepackage{graphicx}

\usepackage{listings}
\begin{document}

\copyrightyear{2025}
\copyrightclause{Copyright for this paper by its authors.
  Use permitted under Creative Commons License Attribution 4.0
  International (CC BY 4.0).}

\conference{SIGIR-e-Com'25: SIGIR Workshop on eCommerce, 2025, Padua, Italy}

\title{NEAR$^2$: A Nested Embedding Approach to Efficient Product Retrieval and Ranking}


\author[1]{Shenbin Qian}[%
email=s.qian@surrey.ac.uk,
]
\author[1]{Diptesh Kanojia}[%
email=d.kanojia@surrey.ac.uk,
]
\cormark[1]
\author[2]{Samarth Agrawal}[%
email=samagrawal@ebay.com,
]
\author[4]{Hadeel Saadany}[%
email=hadeel.saadany@bcu.ac.uk,
]
\author[1]{Swapnil Bhosale}[%
email=s.bhosale@surrey.ac.uk,
]
\author[1]{Constantin Orasan}[%
email=c.orasan@surrey.ac.uk,
]
\author[3]{Zhe Wu}[%
email=zwu1@ebay.com,
]

\address[1]{University of Surrey, United Kingdom}
\address[2]{eBay Inc, Seattle, WA, USA}
\address[3]{eBay Inc, San Jose, CA, USA}
\address[4]{Birmingham City University, United Kingdom}

\cortext[1]{Corresponding author.}

\begin{abstract}
E-commerce information retrieval (IR) systems struggle to simultaneously achieve high accuracy in interpreting complex user queries and maintain efficient processing of vast product catalogs. The dual challenge lies in precisely matching user intent with relevant products while managing the computational demands of real-time search across massive inventories. In this paper, we propose a \textbf{N}ested \textbf{E}mbedding \textbf{A}pproach to product \textbf{R}etrieval and \textbf{R}anking, called NEAR$^2$, which can achieve up to $12$ times efficiency in embedding size at inference time while introducing no extra cost in training and improving performance in accuracy for various encoder-based Transformer models. We validate our approach using different loss functions for the retrieval and ranking task, including multiple negative ranking loss and online contrastive loss, on four different test sets with various IR challenges such as \textit{short and implicit queries}. Our approach achieves an improved performance over a smaller embedding dimension, compared to any existing models.
\end{abstract}


\maketitle

\section{Introduction}

In e-commerce platforms like Amazon, eBay, and Walmart, effective information retrieval (IR) is crucial for matching user queries with relevant products. However, IR systems face dual challenges of accuracy and efficiency. Accurately interpreting the user intent and ranking search results are complicated by ambiguous, repetitive, and alphanumeric queries~\citep{10.1145/3447548.3467101,10.1145/3534965,saadany-etal-2024-centrality}. For example, ``iPhone 13'' often fails to clarify user intent, leading to irrelevant results like ``iPhone 13 case'' being ranked alongside the intended product. Repetition of query terms in both relevant and irrelevant titles exacerbates this issue. For instance, the term ``iPhone 13'' might appear in unrelated accessory titles, confusing embedding-based models. Additionally, alphanumeric queries, such as ``S2716DG'', pose problems because slight variations (\textit{e.g.}, changing ``DG'' to ``DP'') signify different product features, which semantic similarity models struggle to interpret without an exact match. These challenges reflect the difficulty of aligning query interpretation with user intent.

At the same time, the computational demands of processing massive product catalogs in real time make efficient retrieval a pressing concern~\citep{network2040034}. Balancing accuracy with efficiency remains a significant hurdle for modern IR engines. On the efficiency front, current IR systems often rely on computationally intensive models, such as deep neural networks or large-scale embedding computations, to evaluate semantic similarities between queries and product titles~\citep{10184013,Zhu2023-xg}. For instance, calculating embeddings for millions of product titles during a live query can create latency, especially when combined with re-ranking stages that refine results. This latency impacts user experience, as delays of even a fraction of a second can lead to dissatisfaction or abandoned searches. Optimizing these systems to handle large-scale data efficiently without compromising accuracy is a critical challenge in e-commerce search.

In this paper, we propose a \textbf{N}ested \textbf{E}mbedding \textbf{A}pproach to product \textbf{R}etrieval and \textbf{R}anking, called NEAR$^2$, which can achieve efficient product retrieval and ranking using much smaller embedding sizes of encoder-based Transformer models~\citep{Vaswani2017}. This approach maintains performance comparable to the full model without incurring additional training costs. Our evaluation results on various test sets that contain different types of challenging queries, such as implicit and alphanumeric queries, indicate that NEAR$^2$ can improve model performance on these challenging datasets using significantly smaller embedding dimension sizes. Our contributions can be summarized as follows:

\begin{itemize}
\itemsep 0mm

\item We propose NEAR$^2$, a nested embedding approach, which can achieve up to 12$\times$ efficiency in embedding size and 100$\times$ smaller in memory usage during inference while introducing no extra cost in training.
\item We evaluate NEAR$^2$ on four different test sets that contains various types challenging queries. Evaluation results show that our approach achieves an improved performance using a much smaller embedding dimension compared to any existing models.
\item We conduct ablative experiments on different encoder-based models fine-tuned using different IR loss functions. We find that NEAR$^2$ is robust to different IR losses or loss combinations for continued fine-tuning.
\item We perform a qualitative analysis on retrieved product titles using challenging queries. Our analysis re-affirms the superior performance of our approach and reveals that the similarity scores from NEAR$^2$ models are more reliable than those of baseline models.

\end{itemize}

\section{Related Work}

Modern IR systems encounter several challenges that hinder their performance, particularly in dealing with complex queries and data representation. Ambiguities in natural language, vocabulary mismatches, and the need for scalable real-time processing pose significant challenges~\citep{10184013}. Traditional term-based models often fail due to lexical gaps and polysemy, necessitating the transition to advanced semantic models. Semantic retrieval with dense representations, powered by neural networks and pre-trained language models (PTLMs) like BERT~\citep{devlin-etal-2019-bert}, has shown remarkable improvements in handling context and semantics. However, these models demand substantial computational resources and struggle with implicit or alphanumeric queries~\citep{10184013}. Similarly, interaction-based approaches focus on capturing query-document dynamics through deep neural networks, such as the Deep Relevance Matching Model~\citep{10.1145/2983323.2983769}, but often sacrifice efficiency and scalability due to their inability to cache document embeddings offline and their reliance on real-time computation~\citep{10.1145/3038912.3052579}. To gap the mismatch of user intent and retrieved product titles in search queries, \citet{saadany-etal-2024-centrality} curated a dataset annotated with user-intent centrality scores, and proposed a dual loss optimization strategy to fine-tune PTLMs on the dataset in a multi-task learning setting, to solve such challenges. 

To address the efficiency issue, researchers have proposed a range of solutions aimed at enhancing efficiency while maintaining accuracy at the same time. Efficiency issues can be tackled through using DUET models that employ local and distributed deep neural networks, which learns dense lower-dimensional vector representations of the query and the document text for efficient retrieval~\citep{10.1145/3038912.3052579}. Knowledge distillation, where smaller models inherit knowledge from larger PTLMs, has proven effective in reducing resource requirements without compromising performance for IR systems~\citep{kim2023embeddistill}. To mitigate computational overhead, \citet{wan-etal-2022-fast} proposed to use dimension reduction and distilled encoders to create lightweight models for fast and efficient question-answer retrieval. \citet{kusupati2022matryoshka} proposed Matryoshka representation learning (MRL) which is able to encode information at different granularities, to adapt to the computational constraints of various downstream tasks. In this paper, we tackle the challenges of accuracy and efficiency using a nested embedding approach based on MRL to create lightweight embedding models for IR tasks. 

\section{Methodology}

This section describes our nested embedding approach in $\S$~\ref{method_matryo} and the backbone models in $\S$~\ref{backbone_models}.

\subsection{Nested Embedding Training} \label{method_matryo}

We utilize MRL with a ranking loss to train nested embeddings of different sizes on various models.

\begin{figure}[h]
  \centering
  \includegraphics[width=0.7\textwidth]{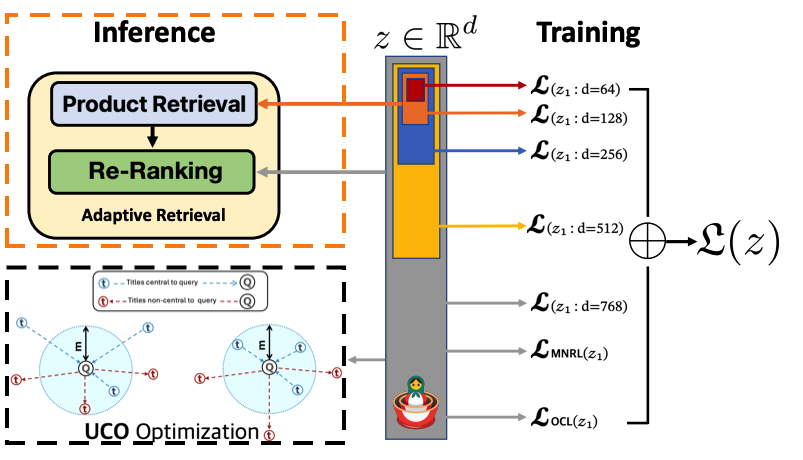}
  \caption{NEAR$^2$ combines UCO optimisation~\cite{saadany-etal-2024-centrality} with MRL~\citep{kusupati2022matryoshka} to learn multiple nested embedding representations of different sizes ($z\in \mathbb{R}^d$ as the full embedding representation) with multi-task learning, which are adaptive to different downstream tasks such as retrieval or ranking during inference.}
  \label{fig.matryoshka}
\end{figure}

\paragraph{Matryoshka Representation Learning} MRL develops representations with diverse capacities within the same higher-dimensional vector by explicitly optimizing sets of lower-dimensional vectors in a nested manner, as illustrated in Figure~\ref{fig.matryoshka}.

The initial $m-$dimensions of the Matryoshka representation, where $m \in M$, the set of nested representation sizes, form a compact and information-dense vector that matches the accuracy of a separately trained $m-$dimensional representation, but requires no extra training effort. As dimensionality increases, the representation progressively incorporates more detailed information, providing a nested coarse-to-fine representation. This approach maintains near-optimal accuracy relative to the full dimensional scale, while avoiding substantial training or deployment costs~\citep{Li2024-re}. 

The MRL loss is formally defined in Equation~\ref{eq:loss_mrl}, where $L_{task}$ is the loss for downstream tasks such as the cross-entropy loss for classification tasks. $f_m(x)$ is the output of the $m$-th nested embedding representation, and $c_m$ is the importance weight for the $m$-th embedding representation. 

\begin{equation} \label{eq:loss_mrl}
    L_{MRL} = \sum_{m \in M}c_m L_{task}(f_m(x), y)
\end{equation}

MRL learns multiple nested embedding representations, each with a different size $m \in M$. The final MRL loss is a weighted sum of the task losses for each of the nested representations. For our product retrieval and ranking task, we set the multiple negative ranking loss (MNRL)~\citep{henderson2017efficient} as our $L_{task}$.

\paragraph{Multiple Negative Ranking Loss} 
MNRL measures the difference between relevant (positive) and irrelevant (negative) examples associated with a given query. This technique ensures a clear separation by reducing the distance between the query and positive samples while increasing the distance from negative samples. Using multiple negative examples enhances the model's ability to discern varying levels of irrelevance, refining its optimization. The MNRL objective function is formulated as follows:

\begin{equation} \label{eq:loss_mnrl}
    \small
    MNRL = \sum_{i=1}^P \sum_{j=1}^N max(0,f(q,p_i) - f(q,n_j) + margin)
\end{equation}

In Equation~\ref{eq:loss_mnrl}, $P$ represents the number of positive samples; $N$ denotes the number of negative samples; $q$ is the query; $f$ is the similarity metric (cosine similarity in our case), and the $margin$ is a hyperparameter defining the ideal distance between positive and negative samples based on the relevance score. The goal of MNRL is to minimize the similarity between $(q,p_i)$ while simultaneously maximizing the difference between $(q,n_j)$ for all positive and negative samples.

\subsection{Backbone Models} \label{backbone_models}

We used encoder-based Transformer models as our backbone for training nested embeddings for efficient product retrieval and ranking. 

\paragraph{Pre-trained Language Models} We initially leveraged BERT~\citep{devlin-etal-2019-bert}, a publicly available pre-trained encoder Transformer model. For our specific use case in e-commerce, we also employed eBERT\footnote{\href{https://innovation.ebayinc.com/tech/engineering/how-ebay-created-a-language-model-with-three-billion-item-titles/}{eBERT Language Model}}, a proprietary multilingual language model pre-trained internally at eBay. This custom model was pre-trained on a corpus of approximately three billion product titles, supplemented by data from general domain sources like Wikipedia and RefinedWeb.

Expanding our experimental approach, we also incorporated eBERT-siam, a fine-tuned variant of eBERT using a Siamese network architecture. This model aims to generate semantically aligned embeddings for item titles, making it particularly effective for similarity-based search and retrieval tasks. Consistent across all models, we maintained a uniform architectural design of $12$ layers with a dimension size of $768$.
 
\paragraph{User-intent Centrality Optimized (UCO) Models} \citet{saadany-etal-2024-centrality, 10.1145/3627673.3679080} show how current IR systems have problems in achieving user-centric product retrieval and ranking due to implicit or alphanumeric queries. They curated a dataset with user-intent centrality scores (see Section~\ref{data_sec}) and proposed a few models optimized for user-intent using an MNRL loss for retrieval and ranking, and an online contrastive loss (OCL) for user-intent centrality. OCL builds on the traditional contrastive loss (CL)  \citep{carlsson2021semantic} approach but introduces a more focused strategy. While conventional CL uses a twin network to evaluate similarities between all data point pairs from the same and different classes, OCL targets only the most challenging and informative pairs within a batch. By prioritizing such cases, OCL refines the loss calculation to focus on the most critical and complex relationships between data points. 

They applied the two losses in a transfer learning setup for eBERT and eBERT-siam models, and performed fine-tuning for centrality classification. Their results indicate that the UCO models achieve an improved performance for retrieval and ranking. Details can be found in~\citet{saadany-etal-2024-centrality}. 

To improve model efficiency and meanwhile leverage optimized performance of the UCO models, we continued training them using NEAR$^2$ for both eBERT-UCO and eBERT-siam-UCO models.

\section{Experimental Setup}

This section explains the datasets we used for training, validating and testing our approach in $\S$~\ref{data_sec}. Implementation details and evaluation metrics are presented in $\S$~\ref{implementation} and $\S$~\ref{eval_metrics} respectively.

\subsection{Data} \label{data_sec}

We utilized eBay's internal graded relevance (IGR) datasets to train our nested embedding representation. These datasets comprise user search queries alongside the product titles retrieved on the platform. They are annotated by humans following specific guidelines to generate two types of buyer-focused relevance labels.

The first is a relevance ranking scheme, where query-title pairs are assigned a rank from (1) Bad, (2) Fair, (3) Good, (4) Excellent, to (5) Perfect. A ``Perfect'' rating signifies an exact match between the query and title, indicating high confidence that the user's needs are fully met, whereas a ``Bad'' rating indicates no alignment between the query and the product title. This ranking methodology aligns with previous studies \cite{10.1145/3326937.3341259,KANG201653}. The second annotation type is a binary centrality score, derived through majority voting among multiple annotators, indicating whether a product aligns with the user's expressed query intent. Centrality scoring differs from relevance ranking in that it assesses whether an item is an outlier or unexpected in the retrieval set versus being a core match to user expectations.

\begin{figure}[h]
    \centering
    \begin{subfigure}[b]{\columnwidth}
        \centering  
        \includegraphics[width=0.7\textwidth]{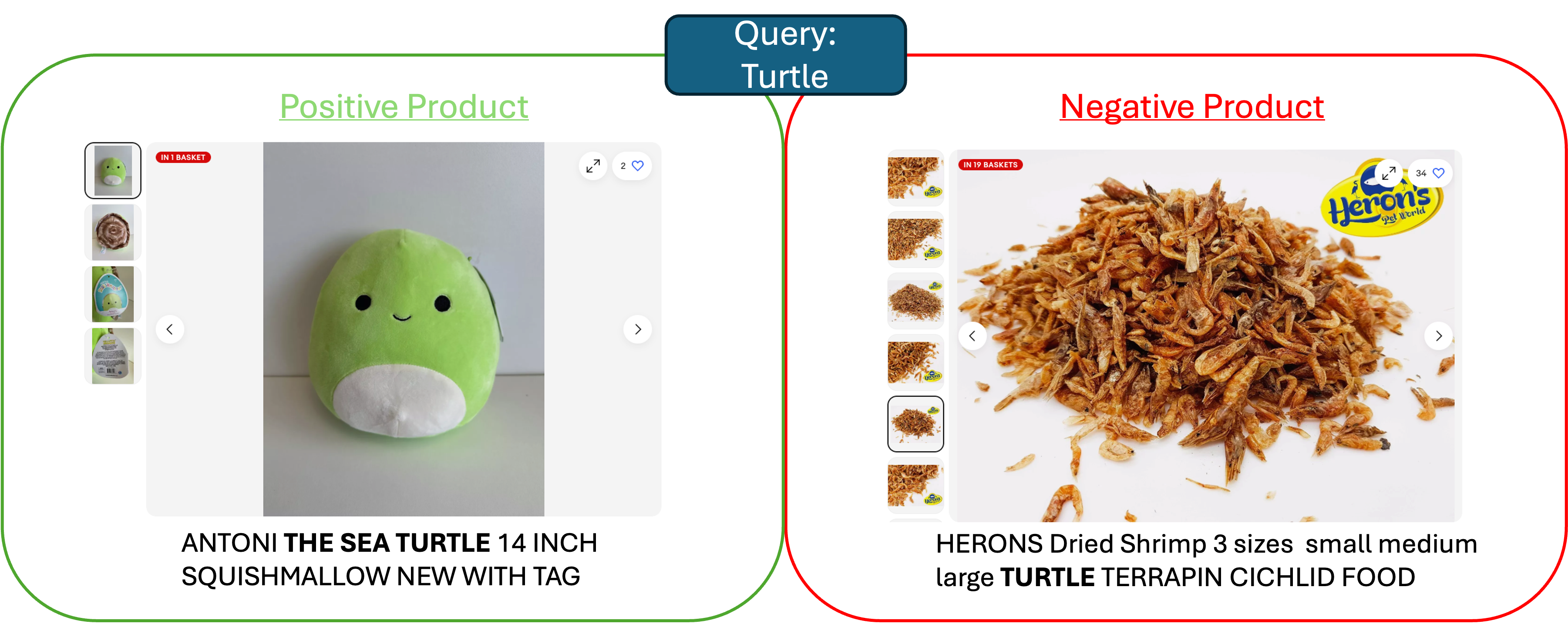}
        \caption{The query ``turtle'' is a part of both positive and negative titles with very different product search outputs. It could also be a part of the ambiguous query ``turtles bepop''.}
        \label{fig:samplesub1}
    \end{subfigure}
    \hfill
    \begin{subfigure}[b]{\columnwidth}
        \centering
        \includegraphics[width=0.7\textwidth]{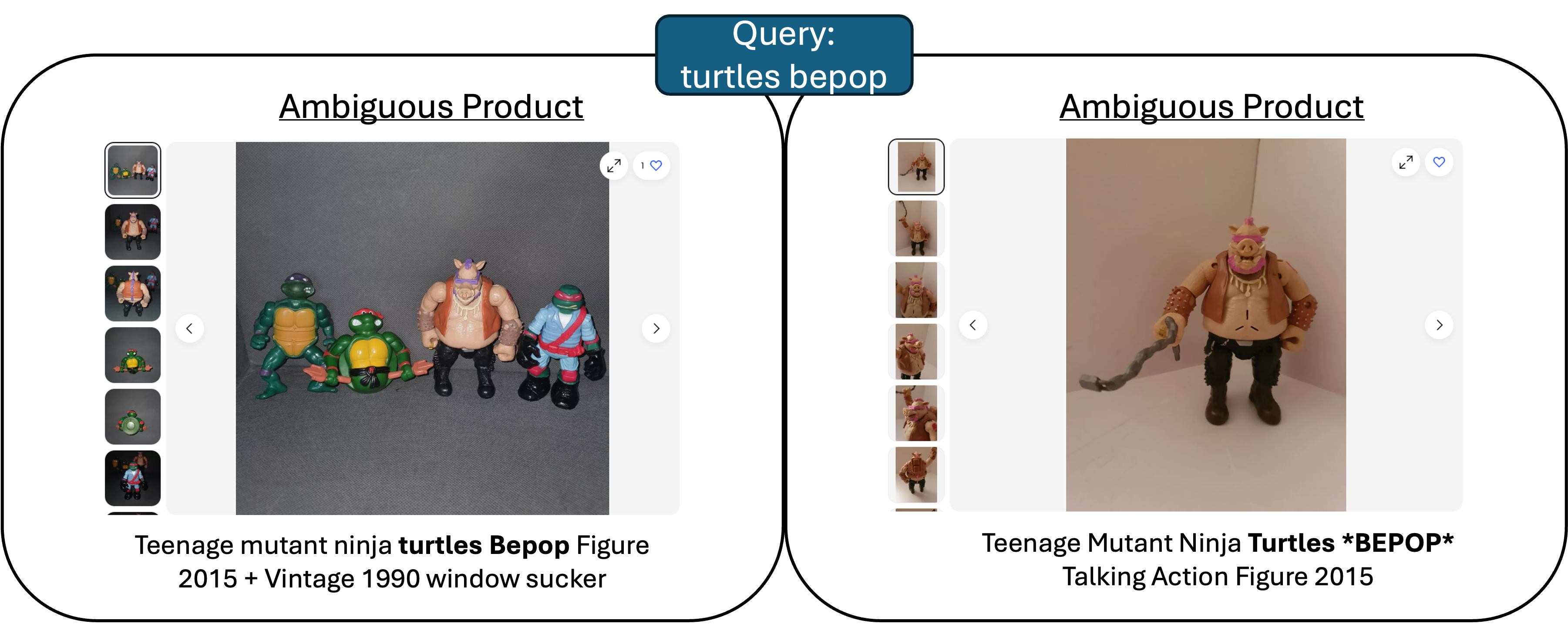}
        \caption{The query ``turtles bepop'' is ambiguous as it could be referred to the major antagonist, ``Bepop'' or together with other Ninjia Turtles.}
        \label{fig:samplesub2}
    \end{subfigure}
    \caption{Examples of query-title pairs from the {\it CQ-common-str} test set. The search queries can be very short and ambiguous, but the retrieved products can be very different as shown in (a), or their titles can be quite close in semantic relation as shown in (b).}
    \label{fig:data_example}
\end{figure}

To compare the results of our approach with those reported in~\citet{saadany-etal-2024-centrality}, we utilized the Common Queries (\textbf{CQ}), CQ Balanced (\textbf{CQ-balanced}), CQ Common String (\textbf{CQ-common-str}), and CQ Alphanumeric (\textbf{CQ-alphanum}) test sets proposed in their paper. The CQ test set was constructed using queries with both positive (relevancy > $3$) and negative (relevancy < $3$) titles, resulting in a dataset skewed toward positive pairs due to the nature of e-commerce data collection. To address this imbalance, a new version, CQ-balanced, was created with approximately equal numbers of positive and negative query-title pairs. The CQ-common-str set was derived by selecting queries where the exact query string appeared in both positive and negative titles, ensuring a strong correlation between relevance scores (both graded relevance and binary centrality). Finally, CQ-alphanum was created to include only query-title pairs containing alphanumeric characters, allowing for a more focused evaluation. Details about their formulation can be found in~\citet{saadany-etal-2024-centrality}. An example of the datasets and the size for each test set can be seen in Figure~\ref{fig:data_example} and Table~\ref{tab:eval_splits}.

\begin{table}[h]
    \centering
    \resizebox{0.5\columnwidth}{!}{%
    \begin{tabular}{c|c|c}
    
    \toprule
    \textbf{Test Name} & \textbf{\# Corpus} & \textbf{\# Queries} \\
    \midrule
    {\it CQ} & $187469$  & $17325$ \\
    {\it CQ-balanced} & $46561$  & $17325$ \\
    {\it CQ-common-str} & $12508$  & $6351$ \\
    {\it CQ-alphanum} & $162115$  & $12333$ \\
    \bottomrule
    \end{tabular}%
    }
    \caption{The size of the four test sets.}
    \label{tab:eval_splits}
\end{table}

\subsection{Implementation Details} \label{implementation}

We continued training the PTLMs and the UCO models in $\S$~\ref{backbone_models} for $2$ epochs, using our nested embedding approach at dimension sizes of $768$, $512$, $256$, $128$ and $64$, on the query-title pairs using only the relevance ranking scores (excluding pairs with a score of $3$) of the IGR datasets. 

During training, we ran a sequential evaluator on the ranking score data to validate for all dimension sizes. First, the evaluator computes the embeddings for both query and title and uses them to calculate the cosine similarity. Then, it finds the most relevant product title to the query (top $3$, $5$ and $10$ titles) in the corpus of all titles with a max corpus size of $200,000$. For all experiments, we set a batch size of $32$, a margin of $0.75$ for the MNRL loss with the AdamW optimizer~\citep{loshchilov2018decoupled} and the learning rate as $5e-05$. Training one model using the above hyperparameters takes $\approx 1.5$ hours on a single NVIDIA V100 GPU.

\subsection{Evaluation Metrics} \label{eval_metrics}

We evaluated the model effectiveness through multiple established evaluation metrics including precision, recall, normalized discounted cumulative gain (NDCG)~\citep{10.1145/582415.582418} and mean reciprocal rank (MRR).

Precision@$k$ quantifies the ratio of pertinent items within the top-$k$ recommended products, focusing on their individual relevance. Conversely, recall@$k$ assesses the proportion of successfully retrieved relevant items compared to the total number of applicable products, regardless of their positioning. NDCG provides a comprehensive assessment of recommendation quality by analyzing both the relevance and positioning of suggested items. This metric compares the actual recommendation order against an idealized ranking, offering a nuanced evaluation of recommendation performance. MRR focuses on measuring the average ranking position of the first relevant item across different queries. A superior MRR indicates the model's capability to prominently feature highly relevant products, thereby enhancing user experience and recommendation effectiveness. 

\section{Results and Discussion}

\begin{table}[h!]
\centering
\resizebox{15cm}{!}{%
\begin{tabular}{ccccccccccc}
\toprule
\multicolumn{1}{c}{\multirow{2}{*}{Model}} & {Precision@$k$} &           &           & Recall@$k$    &           &           & {NDCG@$k$}      &           &           & {MRR@$k$}       \\  
\multicolumn{1}{c}{}                        & 3         & 5         & 10        & 3         & 5         & 10        & 3         & 5         & 10        & 10        \\ \hline
\multicolumn{11}{c}{CQ test}                                                                                                                                        \\ \hline
eBERT-siam                                  & +11.80\%  & +11.79\%  & +11.49\%  & +9.99\%   & +9.72\%   & +9.07\%   & +11.50\%  & +11.23\%  & +10.65\%  & +9.06\%   \\
eBERT-UCO                                   & +2.98\%   & +3.28\%   & +3.90\%   & +3.12\%   & +2.99\%   & +3.16\%   & +3.27\%   & +3.34\%   & +3.47\%   & +3.03\%   \\
eBERT-siam UCO                              & +2.82\%   & +2.75\%   & +3.16\%   & +2.72\%   & +2.45\%   & +2.50\%   & +2.91\%   & +2.77\%   & +2.80\%   & +2.58\%   \\ \hline
\multicolumn{11}{c}{CQ-balanced test}                                                                                                                               \\ \hline
eBERT-siam                                  & +8.85\%   & +8.45\%   & +7.31\%   & +8.85\%   & +8.43\%   & +7.28\%   & +10.28\%  & +10.03\%  & +9.56\%   & +10.48\%  \\
eBERT-UCO                                   & +3.19\%   & +2.87\%   & +2.42\%   & +3.15\%   & +2.81\%   & +2.41\%   & +3.36\%   & +3.19\%   & +3.03\%   & +3.25\%   \\
eBERT-siam UCO                              & +2.77\%   & +2.45\%   & +2.09\%   & +2.75\%   & +2.48\%   & +2.05\%   & +3.06\%   & +2.93\%   & +2.77\%   & +3.01\%   \\ \hline
\multicolumn{11}{c}{CQ-common-str test}                                                                                                                             \\ \hline
eBERT-siam                                  & +6.62\%   & +4.90\%   & +3.00\%   & +6.59\%   & +4.84\%   & +3.01\%   & +8.57\%   & +7.70\%   & +6.99\%   & +8.51\%   \\
eBERT-UCO                                   & +1.69\%   & +1.53\%   & +0.81\%   & +1.68\%   & +1.51\%   & +0.86\%   & +1.56\%   & +1.48\%   & +1.27\%   & +1.38\%   \\
eBERT-siam UCO                              & +1.49\%   & +1.22\%   & +0.81\%   & +1.48\%   & +1.18\%   & +0.83\%   & +1.86\%   & +1.72\%   & +1.59\%   & +1.85\%   \\ \hline
\multicolumn{11}{c}{CQ-alphanum test}     \\ \hline
eBERT-siam                                  & +5.82\%   & +5.84\%   & +6.15\%   & +4.70\%   & +4.59\%   & +5.01\%   & +5.52\%   & +5.40\%   & +5.35\%   & +4.41\%   \\
eBERT-UCO                                   & +3.64\%   & +3.75\%   & +3.92\%   & +3.61\%   & +3.55\%   & +3.60\%   & +3.30\%   & +3.33\%   & +3.40\%   & +2.57\%   \\
eBERT-siam UCO                              & +2.32\%   & +2.13\%   & +2.68\%   & +2.15\%   & +1.87\%   & +2.36\%   & +2.33\%   & +2.13\%   & +2.38\%   & +2.28\%   \\ \hline
\end{tabular}%
}
\caption{Delta in precision, recall, NDCG, and MRR at $k$ on all the test sets for different encoder-based models fine-tuned using  \textbf{NEAR$^2$ at $64$ dimensions} of the entire embedding size ($768$).}
\label{tab:results}
\end{table}
Results achieved using NEAR$^2$ with a dimension size of 64 are shown in Table~\ref{tab:results}. Since BERT and eBERT were not fine-tuned on e-commerce data\footnote{eBERT was only pre-trained on e-commerce data.}, improvement achieved using our approach is huge, as listed in Table~\ref{tab:results_extra} in Appendix~\ref{sec:appendix}. The values are shown as the percentage of increase (delta) of the evaluation metrics in comparison of those without using NEAR$^2$ presented in \citet{saadany-etal-2024-centrality}. 

Comparing results upon using NEAR$^2$ \textit{vs} existing models, we find that our approach remarkably improves performance on all test sets for all models in $\S$~\ref{backbone_models}, even using embeddings with a dimension size of 64, which is $12\times$ smaller in size and more than $100\times$ smaller in memory usage than the full model (see Table~\ref{tab:mem_use}). 

\begin{table}[]
\centering
\resizebox{0.5\columnwidth}{!}{%
\begin{tabular}{cc}
\toprule
Embedding Size & Memory Usage (MB) \\ \hline
768 & 398.03 \\
512 & 2.77 \\
256 & 4.09 \\
128 & 0.55 \\
64 & 1.56 \\
\bottomrule
\end{tabular}%
}
\caption{Memory usage at different embedding sizes for eBERT-siam.}
\label{tab:mem_use}
\end{table}

When comparing results of different dimension sizes from the largest ($768$) to the smallest ($64$), as shown in Table~\ref{tab:CQ_test_dims}\footnote{BERT and eBERT results are in Table~\ref{tab:CQ_test_dims_extra} in Appendix~\ref{sec:appendix}.} for the \textbf{CQ test} set, we discover that the drop in performance is not significant. Embeddings of some smaller dimensions are even slightly better than larger ones. For example, the performance of the eBERT-siam model using NEAR$^2$ at dimension $512$ is slightly better than $768$ for precision, NDCG and MRR. This is also true for other models such as BERT, eBERT and eBERT-UCO, which further indicates the effectiveness of our approach for product retrieval and ranking.

To further validate our approach, we qualitatively compared some product titles retrieved with and without NEAR$^{2}$. The comparison consistently confirmed the superior performance of our method. Full details are presented in Appendix \ref{sec:qa-long}.

\begin{table}[h]
\centering
\resizebox{14cm}{!}{%
\begin{tabular}{cccccc}
\toprule
Model                           & Dimension & Precision@5 & Recall@5 & NDCG@5   & MRR@10   \\ \midrule
\multirow{5}{*}{eBERT-siam}     & 768       & +13.33\%     & +11.77\%  & +13.10\%  & +10.20\%  \\
                                & 512       & +13.35\%     & +11.87\%  & +13.16\%  & +10.30\%  \\
                                & 256       & +13.26\%     & +11.68\%  & +13.05\%  & +10.19\%  \\
                                & 128       & +13.10\%     & +11.37\%  & +12.80\%  & +10.16\%  \\
                                & 64        & +11.79\%     & +9.72\%   & +11.23\%  & +9.06\%   \\ \hline
\multirow{5}{*}{eBERT-UCO}      & 768       & +4.25\%      & +4.04\%   & +4.34\%   & +3.50\%   \\
                                & 512       & +4.27\%      & +3.97\%   & +4.37\%   & +3.57\%   \\
                                & 256       & +4.18\%      & +3.83\%   & +4.23\%   & +3.49\%   \\
                                & 128       & +3.86\%      & +3.52\%   & +3.97\%   & +3.42\%   \\
                                & 64        & +3.28\%      & +2.99\%   & +3.34\%   & +3.03\%   \\ \hline
\multirow{5}{*}{eBERT-siam-UCO} & 768       & +3.85\%      & +3.75\%   & +3.82\%   & +3.05\%   \\
                                & 512       & +3.85\%      & +3.72\%   & +3.81\%   & +3.00\%   \\
                                & 256       & +3.62\%      & +3.47\%   & +3.61\%   & +2.96\%   \\
                                & 128       & +3.46\%      & +3.27\%   & +3.46\%   & +2.96\%   \\
                                & 64        & +2.75\%      & +2.45\%   & +2.77\%   & +2.58\%  \\
                                \bottomrule
\end{tabular}%
}
\caption{Delta in precision, recall, NDCG, and MRR at $k$ on \textbf{CQ test} set for different encoder-based models fine-tuned using \textbf{NEAR$^2$} for all dimension sizes.}
\label{tab:CQ_test_dims}
\end{table}

\section{Ablation Study}

To verify whether continual training using NEAR$^2$ can help improve performance and efficiency when models are initially trained with other losses, we conducted several experiments using eBERT and eBERT-siam for ablation studies. First, we continued training the models using NEAR$^2$, which have been fine-tuned using the MNRL and OCL losses respectively to test if our approach works on each of the two individual losses. Second, we tested training these models using the MRL loss first, and then continued fine-tuning on the MNRL and OCL losses in a multi-task learning setting. The results are contrasted with training without using NEAR$^2$, which are presented as the percentage of increase (delta) in the evaluation metrics in Table~\ref{tab:ablation}. 

\begin{table}[h]
\centering
\resizebox{13cm}{!}{%
\begin{tabular}{ccccc}
\toprule
\multirow{2}{*}{Method} & \multicolumn{2}{c}{eBERT} & \multicolumn{2}{c}{eBERT-siam} \\
                        & NDCG@5      & MRR@10      & NDCG@5         & MRR@10        \\ \midrule
MNRL                    & +4.26\%     & +3.48\%     & +2.98\%        & +2.51\%       \\
OCL                     & +32.09\%    & +22.50\%    & +25.86\%       & +15.66\%      \\
MNRL + OCL              & +3.34\%     & +3.03\%     & +2.77\%        & +2.58\%       \\
MRL: MNRL + OCL         & -3.29\%     & -1.51\%     & -3.26\%        & -1.58\%  \\
\bottomrule
\end{tabular}%
}
\caption{Delta in NDCG@$5$ and MRR@$10$ on the \textbf{CQ test} set for eBERT and eBERT-siam trained using NEAR$^2$ on different loss functions. We continued training these models using \textbf{NEAR$^2$ at $64$ dimensions} of the entire embedding size ($768$) after they were fine-tuned on the MNRL and OCL losses separately or together (MNRL + OCL). We also trained them on the MRL loss first and then on the MNRL and OCL losses (MRL: MNRL + OCL).}
\label{tab:ablation}
\end{table}

Our ablative results suggest that applying the nested embedding approach to training embeddings with lower dimensions can improve performance for all models fine-tuned using the MNRL or OCL losses for retrieval and ranking, with much obvious improvement on the models trained using the OCL loss. However, models trained with the MRL loss first, then fine-tuned using the MNRL and OCL losses, show slight performance degradation in terms of NDCG and MRR. This suggests that our approach is most effective when used after training the model with an IR task loss first.

\section{Conclusion and Future Work}

E-commerce IR systems face the challenge of balancing accurate interpretation of complex user queries with efficient processing of large product catalogs. To address this, we introduced NEAR$^2$, a nested embedding approach for efficient product retrieval and ranking. NEAR$^2$ improves accuracy and achieves up to $12\times$ efficiency in embedding size and 100$\times$ smaller in memory usage during inference, without any increase in pre-training costs. Tested across diverse datasets, including short and implicit queries and alphanumeric queries, our method outperforms existing models with smaller embedding dimensions, demonstrating its robustness across challenging evaluation sets, and with efficiency. Our qualitative analysis reinforces the superior performance of our approach, demonstrating that embeddings generated by NEAR$^2$ models are significantly more reliable than those of baseline models when evaluated based on similarity scores. For future work, we plan to: 1) evaluate our model performance through $A/B$ testing in deployment, 2) leverage internal data to refine larger decoder-based generalist embedding models like NV-embed-v2~\citep{lee2024nvembedimprovedtechniquestraining}, and 3) optimize these models using our NEAR$^2$ approach.

\bibliography{sample-ceur, anthology}

\begin{thebibliography}{22}
\expandafter\ifx\csname natexlab\endcsname\relax\def\natexlab#1{#1}\fi
\providecommand{\url}[1]{\texttt{#1}}
\providecommand{\href}[2]{#2}
\providecommand{\path}[1]{#1}
\providecommand{\DOIprefix}{doi:}
\providecommand{\ArXivprefix}{arXiv:}
\providecommand{\URLprefix}{URL: }
\providecommand{\Pubmedprefix}{pmid:}
\providecommand{\doi}[1]{\href{http://dx.doi.org/#1}{\path{#1}}}
\providecommand{\Pubmed}[1]{\href{pmid:#1}{\path{#1}}}
\providecommand{\bibinfo}[2]{#2}
\ifx\xfnm\relax \def\xfnm[#1]{\unskip,\space#1}\fi
\bibitem[{Li et~al.(2021)Li, Lv, Jin, Lin, Yang, Zeng, Wu, and Ma}]{10.1145/3447548.3467101}
\bibinfo{author}{S.~Li}, \bibinfo{author}{F.~Lv}, \bibinfo{author}{T.~Jin}, \bibinfo{author}{G.~Lin}, \bibinfo{author}{K.~Yang}, \bibinfo{author}{X.~Zeng}, \bibinfo{author}{X.-M. Wu}, \bibinfo{author}{Q.~Ma},
\newblock \bibinfo{title}{Embedding-based product retrieval in taobao search},
\newblock in: \bibinfo{booktitle}{Proceedings of the 27th ACM SIGKDD Conference on Knowledge Discovery \& Data Mining}, KDD '21, \bibinfo{publisher}{Association for Computing Machinery}, \bibinfo{address}{New York, NY, USA}, \bibinfo{year}{2021}, p. \bibinfo{pages}{3181–3189}. \URLprefix \url{https://doi.org/10.1145/3447548.3467101}. \DOIprefix\doi{10.1145/3447548.3467101}.
\bibitem[{Keyvan and Huang(2022)}]{10.1145/3534965}
\bibinfo{author}{K.~Keyvan}, \bibinfo{author}{J.~X. Huang},
\newblock \bibinfo{title}{How to approach ambiguous queries in conversational search: A survey of techniques, approaches, tools, and challenges},
\newblock \bibinfo{journal}{ACM Comput. Surv.} \bibinfo{volume}{55} (\bibinfo{year}{2022}). \URLprefix \url{https://doi.org/10.1145/3534965}. \DOIprefix\doi{10.1145/3534965}.
\bibitem[{Saadany et~al.(2024)Saadany, Bhosale, Agrawal, Kanojia, Orasan, and Wu}]{saadany-etal-2024-centrality}
\bibinfo{author}{H.~Saadany}, \bibinfo{author}{S.~Bhosale}, \bibinfo{author}{S.~Agrawal}, \bibinfo{author}{D.~Kanojia}, \bibinfo{author}{C.~Orasan}, \bibinfo{author}{Z.~Wu},
\newblock \bibinfo{title}{Centrality-aware product retrieval and ranking},
\newblock in: \bibinfo{editor}{F.~Dernoncourt}, \bibinfo{editor}{D.~Preo{\c{t}}iuc-Pietro}, \bibinfo{editor}{A.~Shimorina} (Eds.), \bibinfo{booktitle}{Proceedings of the 2024 Conference on Empirical Methods in Natural Language Processing: Industry Track}, \bibinfo{publisher}{Association for Computational Linguistics}, \bibinfo{address}{Miami, Florida, US}, \bibinfo{year}{2024}, pp. \bibinfo{pages}{215--224}. \URLprefix \url{https://aclanthology.org/2024.emnlp-industry.17}.
\bibitem[{Mhawi et~al.(2022)Mhawi, Oleiwi, Saeed, and Al-Taie}]{network2040034}
\bibinfo{author}{D.~N. Mhawi}, \bibinfo{author}{H.~W. Oleiwi}, \bibinfo{author}{N.~H. Saeed}, \bibinfo{author}{H.~L. Al-Taie},
\newblock \bibinfo{title}{An efficient information retrieval system using evolutionary algorithms},
\newblock \bibinfo{journal}{Network} \bibinfo{volume}{2} (\bibinfo{year}{2022}) \bibinfo{pages}{583--605}. \URLprefix \url{https://www.mdpi.com/2673-8732/2/4/34}. \DOIprefix\doi{10.3390/network2040034}.
\bibitem[{Hambarde and Proença(2023)}]{10184013}
\bibinfo{author}{K.~A. Hambarde}, \bibinfo{author}{H.~Proença},
\newblock \bibinfo{title}{Information retrieval: Recent advances and beyond},
\newblock \bibinfo{journal}{IEEE Access} \bibinfo{volume}{11} (\bibinfo{year}{2023}) \bibinfo{pages}{76581--76604}. \DOIprefix\doi{10.1109/ACCESS.2023.3295776}.
\bibitem[{Zhu et~al.(2023)Zhu, Yuan, Wang, Liu, Liu, Deng, Chen, Liu, Dou, and Wen}]{Zhu2023-xg}
\bibinfo{author}{Y.~Zhu}, \bibinfo{author}{H.~Yuan}, \bibinfo{author}{S.~Wang}, \bibinfo{author}{J.~Liu}, \bibinfo{author}{W.~Liu}, \bibinfo{author}{C.~Deng}, \bibinfo{author}{H.~Chen}, \bibinfo{author}{Z.~Liu}, \bibinfo{author}{Z.~Dou}, \bibinfo{author}{J.-R. Wen},
\newblock \bibinfo{title}{Large language models for information retrieval: A survey},
\newblock \bibinfo{journal}{arXiv preprint}  (\bibinfo{year}{2023}). \href{http://arxiv.org/abs/2308.07107}{{\tt arXiv:2308.07107}}.
\bibitem[{Vaswani et~al.(2017)Vaswani, Shazeer, Parmar, Uszkoreit, Jones, Gomez, Kaiser, and Polosukhin}]{Vaswani2017}
\bibinfo{author}{A.~Vaswani}, \bibinfo{author}{N.~Shazeer}, \bibinfo{author}{N.~Parmar}, \bibinfo{author}{J.~Uszkoreit}, \bibinfo{author}{L.~Jones}, \bibinfo{author}{A.~N. Gomez}, \bibinfo{author}{L.~Kaiser}, \bibinfo{author}{I.~Polosukhin},
\newblock \bibinfo{title}{Attention is all you need},
\newblock in: \bibinfo{booktitle}{Proceedings of the 31st International Conference on Neural Information Processing Systems}, NIPS'17, \bibinfo{publisher}{Curran Associates Inc.}, \bibinfo{address}{Red Hook, NY, USA}, \bibinfo{year}{2017}, p. \bibinfo{pages}{6000–6010}.
\bibitem[{Devlin et~al.(2019)Devlin, Chang, Lee, and Toutanova}]{devlin-etal-2019-bert}
\bibinfo{author}{J.~Devlin}, \bibinfo{author}{M.-W. Chang}, \bibinfo{author}{K.~Lee}, \bibinfo{author}{K.~Toutanova},
\newblock \bibinfo{title}{{BERT}: Pre-training of deep bidirectional transformers for language understanding},
\newblock in: \bibinfo{editor}{J.~Burstein}, \bibinfo{editor}{C.~Doran}, \bibinfo{editor}{T.~Solorio} (Eds.), \bibinfo{booktitle}{Proceedings of the 2019 Conference of the North {A}merican Chapter of the Association for Computational Linguistics: Human Language Technologies, Volume 1 (Long and Short Papers)}, \bibinfo{publisher}{Association for Computational Linguistics}, \bibinfo{address}{Minneapolis, Minnesota}, \bibinfo{year}{2019}, pp. \bibinfo{pages}{4171--4186}. \URLprefix \url{https://aclanthology.org/N19-1423}. \DOIprefix\doi{10.18653/v1/N19-1423}.
\bibitem[{Guo et~al.(2016)Guo, Fan, Ai, and Croft}]{10.1145/2983323.2983769}
\bibinfo{author}{J.~Guo}, \bibinfo{author}{Y.~Fan}, \bibinfo{author}{Q.~Ai}, \bibinfo{author}{W.~B. Croft},
\newblock \bibinfo{title}{A deep relevance matching model for ad-hoc retrieval},
\newblock in: \bibinfo{booktitle}{Proceedings of the 25th ACM International on Conference on Information and Knowledge Management}, CIKM '16, \bibinfo{publisher}{Association for Computing Machinery}, \bibinfo{address}{New York, NY, USA}, \bibinfo{year}{2016}, p. \bibinfo{pages}{55–64}. \URLprefix \url{https://doi.org/10.1145/2983323.2983769}. \DOIprefix\doi{10.1145/2983323.2983769}.
\bibitem[{Mitra et~al.(2017)Mitra, Diaz, and Craswell}]{10.1145/3038912.3052579}
\bibinfo{author}{B.~Mitra}, \bibinfo{author}{F.~Diaz}, \bibinfo{author}{N.~Craswell},
\newblock \bibinfo{title}{Learning to match using local and distributed representations of text for web search},
\newblock in: \bibinfo{booktitle}{Proceedings of the 26th International Conference on World Wide Web}, WWW '17, \bibinfo{publisher}{International World Wide Web Conferences Steering Committee}, \bibinfo{address}{Republic and Canton of Geneva, CHE}, \bibinfo{year}{2017}, p. \bibinfo{pages}{1291–1299}. \URLprefix \url{https://doi.org/10.1145/3038912.3052579}. \DOIprefix\doi{10.1145/3038912.3052579}.
\bibitem[{Kim et~al.(2023)Kim, Rawat, Zaheer, Jayasumana, Sadhanala, Jitkrittum, Menon, Fergus, and Kumar}]{kim2023embeddistill}
\bibinfo{author}{S.~Kim}, \bibinfo{author}{A.~S. Rawat}, \bibinfo{author}{M.~Zaheer}, \bibinfo{author}{S.~Jayasumana}, \bibinfo{author}{V.~Sadhanala}, \bibinfo{author}{W.~Jitkrittum}, \bibinfo{author}{A.~K. Menon}, \bibinfo{author}{R.~Fergus}, \bibinfo{author}{S.~Kumar}, \bibinfo{title}{Embeddistill: A geometric knowledge distillation for information retrieval}, \bibinfo{year}{2023}. \URLprefix \url{https://openreview.net/forum?id=BT03V9Re9a}.
\bibitem[{Wan et~al.(2022)Wan, Patel, Murdock, Potdar, and Joshi}]{wan-etal-2022-fast}
\bibinfo{author}{H.~Wan}, \bibinfo{author}{S.~S. Patel}, \bibinfo{author}{J.~W. Murdock}, \bibinfo{author}{S.~Potdar}, \bibinfo{author}{S.~Joshi},
\newblock \bibinfo{title}{Fast and light-weight answer text retrieval in dialogue systems},
\newblock in: \bibinfo{editor}{A.~Loukina}, \bibinfo{editor}{R.~Gangadharaiah}, \bibinfo{editor}{B.~Min} (Eds.), \bibinfo{booktitle}{Proceedings of the 2022 Conference of the North American Chapter of the Association for Computational Linguistics: Human Language Technologies: Industry Track}, \bibinfo{publisher}{Association for Computational Linguistics}, \bibinfo{address}{Hybrid: Seattle, Washington + Online}, \bibinfo{year}{2022}, pp. \bibinfo{pages}{334--343}. \URLprefix \url{https://aclanthology.org/2022.naacl-industry.37}. \DOIprefix\doi{10.18653/v1/2022.naacl-industry.37}.
\bibitem[{Kusupati et~al.(2022)Kusupati, Bhatt, Rege, Wallingford, Sinha, Ramanujan, Howard-Snyder, Chen, Kakade, Jain et~al.}]{kusupati2022matryoshka}
\bibinfo{author}{A.~Kusupati}, \bibinfo{author}{G.~Bhatt}, \bibinfo{author}{A.~Rege}, \bibinfo{author}{M.~Wallingford}, \bibinfo{author}{A.~Sinha}, \bibinfo{author}{V.~Ramanujan}, \bibinfo{author}{W.~Howard-Snyder}, \bibinfo{author}{K.~Chen}, \bibinfo{author}{S.~Kakade}, \bibinfo{author}{P.~Jain}, et~al.,
\newblock \bibinfo{title}{Matryoshka representation learning},
\newblock in: \bibinfo{booktitle}{Advances in Neural Information Processing Systems}, \bibinfo{year}{2022}.
\bibitem[{Li et~al.(2024)Li, Li, Li, Xie, and Li}]{Li2024-re}
\bibinfo{author}{X.~Li}, \bibinfo{author}{Z.~Li}, \bibinfo{author}{J.~Li}, \bibinfo{author}{H.~Xie}, \bibinfo{author}{Q.~Li},
\newblock \bibinfo{title}{{ESE}: Espresso sentence embeddings},
\newblock \bibinfo{journal}{arXiv preprint}  (\bibinfo{year}{2024}). \href{http://arxiv.org/abs/2402.14776}{{\tt arXiv:2402.14776}}.
\bibitem[{Henderson et~al.(2017)Henderson, Al-Rfou, Strope, Sung, Luk{\'a}cs, Guo, Kumar, Miklos, and Kurzweil}]{henderson2017efficient}
\bibinfo{author}{M.~Henderson}, \bibinfo{author}{R.~Al-Rfou}, \bibinfo{author}{B.~Strope}, \bibinfo{author}{Y.-H. Sung}, \bibinfo{author}{L.~Luk{\'a}cs}, \bibinfo{author}{R.~Guo}, \bibinfo{author}{S.~Kumar}, \bibinfo{author}{B.~Miklos}, \bibinfo{author}{R.~Kurzweil},
\newblock \bibinfo{title}{Efficient natural language response suggestion for smart reply},
\newblock \bibinfo{journal}{arXiv preprint arXiv:1705.00652}  (\bibinfo{year}{2017}).
\bibitem[{Saadany et~al.(2024)Saadany, Bhosale, Agrawal, Wu, Or\u{a}san, and Kanojia}]{10.1145/3627673.3679080}
\bibinfo{author}{H.~Saadany}, \bibinfo{author}{S.~Bhosale}, \bibinfo{author}{S.~Agrawal}, \bibinfo{author}{Z.~Wu}, \bibinfo{author}{C.~Or\u{a}san}, \bibinfo{author}{D.~Kanojia},
\newblock \bibinfo{title}{Product retrieval and ranking for alphanumeric queries},
\newblock in: \bibinfo{booktitle}{Proceedings of the 33rd ACM International Conference on Information and Knowledge Management}, CIKM '24, \bibinfo{publisher}{Association for Computing Machinery}, \bibinfo{address}{New York, NY, USA}, \bibinfo{year}{2024}, p. \bibinfo{pages}{5564–5565}. \URLprefix \url{https://doi.org/10.1145/3627673.3679080}. \DOIprefix\doi{10.1145/3627673.3679080}.
\bibitem[{Carlsson et~al.(2021)Carlsson, Gyllensten, Gogoulou, Hellqvist, and Sahlgren}]{carlsson2021semantic}
\bibinfo{author}{F.~Carlsson}, \bibinfo{author}{A.~C. Gyllensten}, \bibinfo{author}{E.~Gogoulou}, \bibinfo{author}{E.~Y. Hellqvist}, \bibinfo{author}{M.~Sahlgren},
\newblock \bibinfo{title}{Semantic re-tuning with contrastive tension},
\newblock in: \bibinfo{booktitle}{International Conference on Learning Representations}, \bibinfo{year}{2021}. \URLprefix \url{https://openreview.net/forum?id=Ov_sMNau-PF}.
\bibitem[{Jiang et~al.(2019)Jiang, Shang, Li, Yang, Tang, Ma, Xiao, and Zhao}]{10.1145/3326937.3341259}
\bibinfo{author}{Y.~Jiang}, \bibinfo{author}{Y.~Shang}, \bibinfo{author}{R.~Li}, \bibinfo{author}{W.-Y. Yang}, \bibinfo{author}{G.~Tang}, \bibinfo{author}{C.~Ma}, \bibinfo{author}{Y.~Xiao}, \bibinfo{author}{E.~Zhao},
\newblock \bibinfo{title}{A unified neural network approach to e-commerce relevance learning},
\newblock in: \bibinfo{booktitle}{Proceedings of the 1st International Workshop on Deep Learning Practice for High-Dimensional Sparse Data}, DLP-KDD '19, \bibinfo{publisher}{Association for Computing Machinery}, \bibinfo{address}{New York, NY, USA}, \bibinfo{year}{2019}. \URLprefix \url{https://doi.org/10.1145/3326937.3341259}. \DOIprefix\doi{10.1145/3326937.3341259}.
\bibitem[{Kang et~al.(2016)Kang, Jang, and Park}]{KANG201653}
\bibinfo{author}{D.~Kang}, \bibinfo{author}{W.~Jang}, \bibinfo{author}{Y.~Park},
\newblock \bibinfo{title}{Evaluation of e-commerce websites using fuzzy hierarchical topsis based on e-s-qual},
\newblock \bibinfo{journal}{Applied Soft Computing} \bibinfo{volume}{42} (\bibinfo{year}{2016}) \bibinfo{pages}{53--65}. \URLprefix \url{https://www.sciencedirect.com/science/article/pii/S1568494616300047}. \DOIprefix\doi{https://doi.org/10.1016/j.asoc.2016.01.017}.
\bibitem[{Loshchilov and Hutter(2019)}]{loshchilov2018decoupled}
\bibinfo{author}{I.~Loshchilov}, \bibinfo{author}{F.~Hutter},
\newblock \bibinfo{title}{Decoupled weight decay regularization},
\newblock in: \bibinfo{booktitle}{International Conference on Learning Representations}, \bibinfo{year}{2019}. \URLprefix \url{https://openreview.net/forum?id=Bkg6RiCqY7}.
\bibitem[{J\"{a}rvelin and Kek\"{a}l\"{a}inen(2002)}]{10.1145/582415.582418}
\bibinfo{author}{K.~J\"{a}rvelin}, \bibinfo{author}{J.~Kek\"{a}l\"{a}inen},
\newblock \bibinfo{title}{Cumulated gain-based evaluation of ir techniques},
\newblock \bibinfo{journal}{ACM Trans. Inf. Syst.} \bibinfo{volume}{20} (\bibinfo{year}{2002}) \bibinfo{pages}{422–446}. \URLprefix \url{https://doi.org/10.1145/582415.582418}. \DOIprefix\doi{10.1145/582415.582418}.
\bibitem[{Lee et~al.(2024)Lee, Roy, Xu, Raiman, Shoeybi, Catanzaro, and Ping}]{lee2024nvembedimprovedtechniquestraining}
\bibinfo{author}{C.~Lee}, \bibinfo{author}{R.~Roy}, \bibinfo{author}{M.~Xu}, \bibinfo{author}{J.~Raiman}, \bibinfo{author}{M.~Shoeybi}, \bibinfo{author}{B.~Catanzaro}, \bibinfo{author}{W.~Ping}, \bibinfo{title}{Nv-embed: Improved techniques for training llms as generalist embedding models}, \bibinfo{year}{2024}. \URLprefix \url{https://arxiv.org/abs/2405.17428}. \href{http://arxiv.org/abs/2405.17428}{{\tt arXiv:2405.17428}}.

\end{thebibliography}

\clearpage

\appendix

\section{Additional Figures and Tables} \label{sec:appendix}

\counterwithin{figure}{section}
\setcounter{figure}{0} 
\counterwithin{table}{section}
\setcounter{table}{0} 

\begin{table}[h]
\centering
\resizebox{16cm}{!}{%
\begin{tabular}{ccccccccccc}
\toprule
\multicolumn{1}{c}{\multirow{2}{*}{Model}} & {Precision@$k$} &           &           & Recall@$k$    &           &           & {NDCG@$k$}      &           &           & {MRR@$k$}       \\  
\multicolumn{1}{c}{}                        & 3         & 5         & 10        & 3         & 5         & 10        & 3         & 5         & 10        & 10        \\ \hline
\multicolumn{11}{c}{CQ test}                                                                                                                                        \\ \hline
BERT                                        & +244.88\% & +274.90\% & +296.75\% & +261.89\% & +278.42\% & +277.64\% & +230.75\% & +251.93\% & +263.34\% & +164.96\% \\
eBERT                                       & +185.18\% & +198.72\% & +196.57\% & +204.36\% & +202.87\% & +185.80\% & +180.69\% & +191.63\% & +190.49\% & +124.19\%   \\ \hline
\multicolumn{11}{c}{CQ-balanced test}                                                                                                                               \\ \hline
BERT                                        & +261.57\% & +239.27\% & +207.46\% & +262.23\% & +239.63\% & +207.94\% & +273.74\% & +261.24\% & +245.48\% & +262.73\% \\
eBERT                                       & +178.60\% & +151.73\% & +121.33\% & +178.84\% & +151.59\% & +121.06\% & +197.78\% & +181.60\% & +164.85\% & +186.77\%  \\ \hline
\multicolumn{11}{c}{CQ-common-str test}                                                                                                                             \\ \hline
BERT                                        & +230.82\% & +206.66\% & +171.23\% & +230.87\% & +206.58\% & +171.61\% & +238.59\% & +226.38\% & +210.68\% & +226.68\% \\
eBERT                                       & +148.89\% & +125.23\% & +98.80\%  & +148.64\% & +125.10\% & +98.64\%  & +167.57\% & +154.43\% & +141.09\% & +160.48\%  \\ \hline
\multicolumn{11}{c}{CQ-alphanum test}     \\ \hline
BERT                                        & +176.19\% & +202.04\% & +215.87\% & +177.55\% & +199.48\% & +201.58\% & +164.68\% & +181.94\% & +188.90\% & +117.16\% \\
eBERT                                       & +160.04\% & +176.97\% & +181.05\% & +161.56\% & +170.52\% & +165.06\% & +152.21\% & +163.12\% & +164.35\% & +104.15\% \\ \hline
\end{tabular}%
}
\caption{Delta in precision, recall, NDCG, and MRR at $k$ on all the test sets for BERT and eBERT fine-tuned using  \textbf{NEAR$^2$ at $64$ dimensions} of the entire embedding size ($768$).}
\label{tab:results_extra}
\end{table}

\begin{table}[h]
\centering
\resizebox{12cm}{!}{%
\begin{tabular}{cccccc}
\toprule
Model                           & Dimension & Precision@5 & Recall@5 & NDCG@5   & MRR@10   \\ \midrule
 \multirow{5}{*}{BERT}           & 768       & +286.80\%    & +296.32\% & +265.40\% & +170.99\% \\
                                 & 512       & +287.11\%    & +296.11\% & +265.57\% & +171.13\% \\
                                 & 256       & +286.80\%    & +295.49\% & +264.91\% & +170.52\% \\
                                 & 128       & +284.27\%    & +291.95\% & +262.16\% & +169.54\% \\
                                 & 64        & +274.90\%    & +278.42\% & +251.93\% & +164.96\% \\ \hline
 \multirow{5}{*}{eBERT}          & 768       & +192.17\%    & +197.60\% & +185.11\% & +119.88\% \\
                                 & 512       & +192.41\%    & +197.45\% & +185.24\% & +120.00\% \\
                                 & 256       & +192.17\%    & +196.98\% & +184.72\% & +119.50\% \\
                                 & 128       & +190.26\%    & +194.32\% & +182.58\% & +118.71\% \\
                                 & 64        & +183.19\%    & +184.16\% & +174.59\% & +114.99\% \\ 
                                \bottomrule
\end{tabular}%
}
\caption{Delta in precision, recall, NDCG, and MRR at $k$ on \textbf{CQ test} set for BERT and eBERT models fine-tuned using \textbf{NEAR$^2$} for all dimension sizes.}
\label{tab:CQ_test_dims_extra}
\end{table}

\section{Detailed Qualitative Analysis} \label{sec:qa-long}

To understand the performance improvements of our approach compared to existing models, we conducted a qualitative analysis using examples from the \textbf{CQ test} set. Specifically, we generated inferences for all instances in the CQ test set with eBERT and eBERT-siam\footnote{We mainly analyze results from eBERT. Results from eBERT-siam can be seen in Tables~\ref{tab:error_example1siam} and~\ref{tab:error_example2siam}.} using or not using the NEAR$^2$ approach at a dimension size of 64 (NEAR$^2$@64). For each query, we retrieved the top 10 product titles and ranked them based on their cosine similarity scores. To evaluate real-world performance, we selected two representative queries: one short and implicit query and one long and detailed query. These examples provided insights into how our approach performs relative to eBERT or eBERT-siam in practical scenarios. 

\begin{table*}[h]
\centering
\resizebox{15cm}{!}{%
\begin{tabular}{cccc}
\toprule
Method                         & Retrieved Title                                                                    & Ranking & Sim\_Score$_{Norm}$ \\ \midrule
\multirow{10}{*}{NEAR$^2$@64} & Philodendron Micans Rooted Cutting Trailing House Plant Cuttings Rare   Plants     & 1       & 0.3935         \\
                              & Tillandsia Mix  5 Plants    Indoor Air Plant for House Vivarium Terrarium          & 2       & 0.3880          \\
                              & Big leaf philodendron pink princess plant   cutting  1 leaf cutting                & 3       & 0.3760         \\
                              & 2 NEON PINK SALVIA PLANT  PERENNIAL SAGE HIGHLY FRAGRANT                           & 4       & 0.3725         \\
                              & Spathiphyllum Peace Lily Indoor   Plants  1 x Potted Lily House Plant 9cm   Pot    & 5       & 0.3693         \\
                              & Cissus Discolor aka Rex Begonia Vine 6   inch pot                                  & 6       & 0.3687         \\
                              & 3 Plant    4 Pots  Great Houseplant   Assorted Rex Begonia  Easy to grow   housepl & 7       & 0.3684         \\
                              & PHILODENDRON MELANOCHRYSUM VERY LARGE   25  3 FEET TALL STUNNING PLANT             & 8       & 0.3679         \\
                              & Spathiphyllum Peace Lily House Plant  Live Indoor House Potted Tree In 9cm         & 9       & 0.3620         \\
                              & PHILODENDRON PINK PRINCESS  LARGE    PLANT IN 15CM POT HOUSE PLANT                 & 10      & 0.3593         \\ \hline
\multirow{10}{*}{W/o NEAR$^2$@64}       & Avocado plant                                                                      & 1       & 0.0604         \\
                              & coins                                                                              & 2       & 0.0520         \\
                              & Begonia    Butterfly                                                               & 3       & 0.0494         \\
                              & drinks cabinet                                                                     & 4       & 0.0487         \\
                              & Eucalyptus tree                                                                    & 5       & 0.0483         \\
                              & portfolio landscape lights                                                         & 6       & 0.0479         \\
                              & Nico    the marble index                                                           & 7       & 0.0469         \\
                              & car assessories                                                                    & 8       & 0.0468         \\
                              & Begonia Curly Q                                                                    & 9       & 0.0465         \\
                              & Houseplant and Pot  Package                                                        & 10      & 0.0454          \\ \hline
Gold label                    & Aloe Vera Plant - Large Plant in Pot                                               & /       & /           \\ \bottomrule    
\end{tabular}%
}
\caption{Retrieved titles for the short and implicit query ``\textbf{plants}'' using or not using NEAR$^2$@64 on \textbf{eBERT}.}
\label{tab:error_example1}
\end{table*}

\paragraph{Short and Implicit Query}
Table~\ref{tab:error_example1} illustrates the retrieved titles, their rankings (from 1 to 10), and their normalized\footnote{Against the minimum value.} similarity scores for the short and implicit query ``plants'' with eBERT. Based on the gold label, the expected product title should include ``potted plants''. For the model using NEAR$^2$@64, all retrieved product titles contained relevant keywords such as ``plant'' or ``pot'', along with detailed product descriptions. In contrast, the titles retrieved by the model without using NEAR$^2$@64 were significantly shorter, with many lacking the keyword ``plant'' and some, such as ``coins'', being entirely irrelevant to the query. Notably, the normalized similarity scores from without using NEAR$^2$@64 are much lower than those of using NEAR$^2$@64, which is responsible for those irrelevant titles retrieved. This highlights the unreliability of the similarity scores from models without using NEAR$^2$.

\begin{table*}[h]
\centering
\resizebox{15cm}{!}{%
\begin{tabular}{cccc}
\toprule
Method                         & Retrieved Title                                                                    & Ranking & Sim\_Score$_{Norm}$ \\ \midrule
\multirow{10}{*}{NEAR$^2$@64} & Vintage 925 Sterling   Silver Fiery Boulder  Opal Ring Uk Size   P                 & 1       & 0.3561         \\
                              & Sterling Silver 925 Signed Opal Heart   Pendant Necklace 19 Chain                  & 2       & 0.3406         \\
                              & Vintage Possibly Opal Pendant On Gold   Tone Necklace Chain                        & 3       & 0.3376         \\
                              & Ethiopian OPAL 083 carat sterling silver   solitaire pendant                       & 4       & 0.3373         \\
                              & Vintage Ring White Opal Fire Lustre Genuine   Natural Gems Sterling Silver Size L  & 5       & 0.3347         \\
                              & Australian Triplet Opal Gemstone 925   Sterling Silver Handmade Ring All Size      & 6       & 0.3342         \\
                              & Moonstone Opal Pendant 925 Sterling   Silver Necklace Earring Women Jewellery Gift & 7       & 0.3314         \\
                              & Green Triplet Fire Opal Peridot 925   Sterling Silver Jewelry Pendants 27 v957     & 8       & 0.3283         \\
                              & Gemporia Mosaic Opal  White Topaz Sterling Silver Pendant Aggl98                   & 9       & 0.3144         \\
                              & Coober Pedy Semi Black Opal pendant 094   carats 179 grams of 925 Sterling Sil     & 10      & 0.3125         \\ \hline
\multirow{10}{*}{W/o NEAR$^2$@64}       & GENUINE 9ct gold gf garnet hoop   earringsPacked full of dazzling stones 7b Y64 7d & 1       & 0.0973         \\
                              & GENUINE 9ct gold gf garnet hoop   earringsPacked full of gemstones 7b Y64 7d       & 2       & 0.0965          \\
                              & Large Vintage Sterling silver cabochon   amethyst  garnet pendant  chain 155g      & 3       & 0.0929         \\
                              & Gold diamante encrusted large round pendant   80 cm long chain rope necklace       & 4       & 0.0925         \\
                              & Vintage 70s sterling silver and oval   amethyst pendant and 925 chain necklace     & 5       & 0.0919         \\
                              & 9ct yellow gold reversible small crystal   puffy love heart pull through earrings  & 6       & 0.0918         \\
                              & CLASSIC 9ct Gold gf Aquamarine hoop   earringsTRULY STUNNING EARRINGS 7b J067 7d   & 7       & 0.0917         \\
                              & STUNNING 9ct Gold Opal toe ring gf WHILE   STOCKS LAST DONT MISS 7b TO88 7d        & 8       & 0.0915         \\
                              & Vintage Art Nouveau style sterling silver   925 and onyx stone scroll leaf brooch  & 9       & 0.0914         \\
                              & Brand new set of two pair of earrings one   butterfly one little girlin a gift box & 10      & 0.0911         \\ \hline
Gold label                    & 925 Sterling Silver Red Coral Gemstone   Handmade Jewelry Vintage Pendant S120     & /       & /              \\ \bottomrule 
\end{tabular}%
}
\caption{Retrieved titles for the long and detailed query ``\textbf{925 sterling silver triplet opal gemstone jewelry vintage pendant s-1.20}'' using or not using NEAR$^2$@64 on \textbf{eBERT}.}
\label{tab:error_example2}
\end{table*}

\begin{figure}[h]
    \centering
    \begin{subfigure}[b]{\columnwidth}
        \centering
        \includegraphics[width=\textwidth]{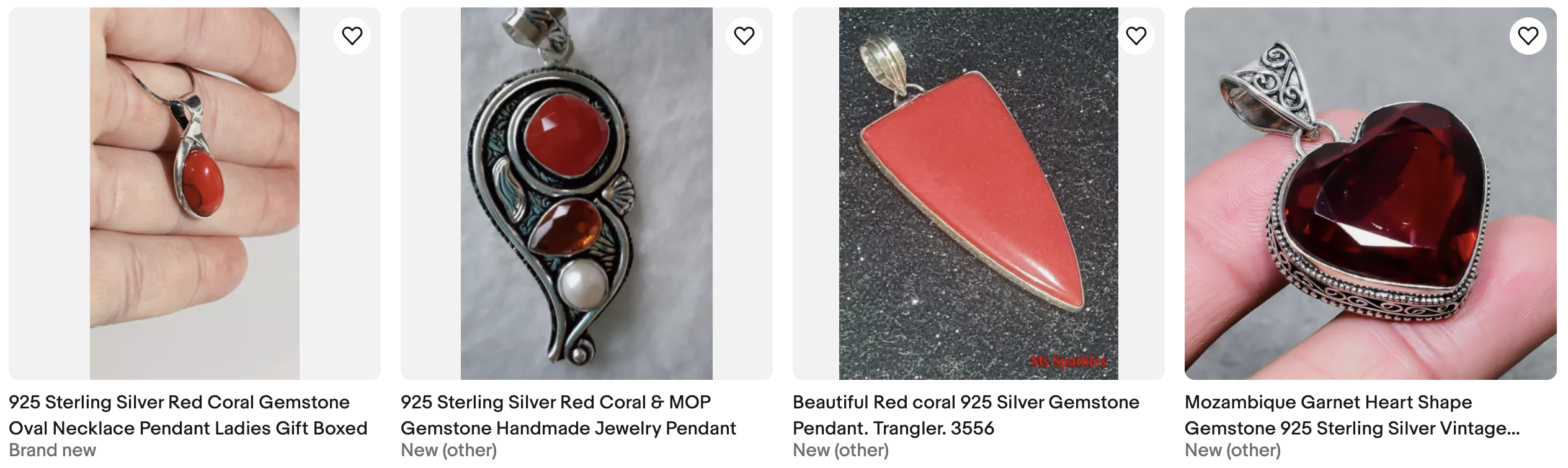}
        \caption{Products retrieved using the gold label title.}
        \label{fig:errorsub1}
    \end{subfigure}
    \hfill
    \begin{subfigure}[b]{\columnwidth}
        \centering
        \includegraphics[width=\textwidth]{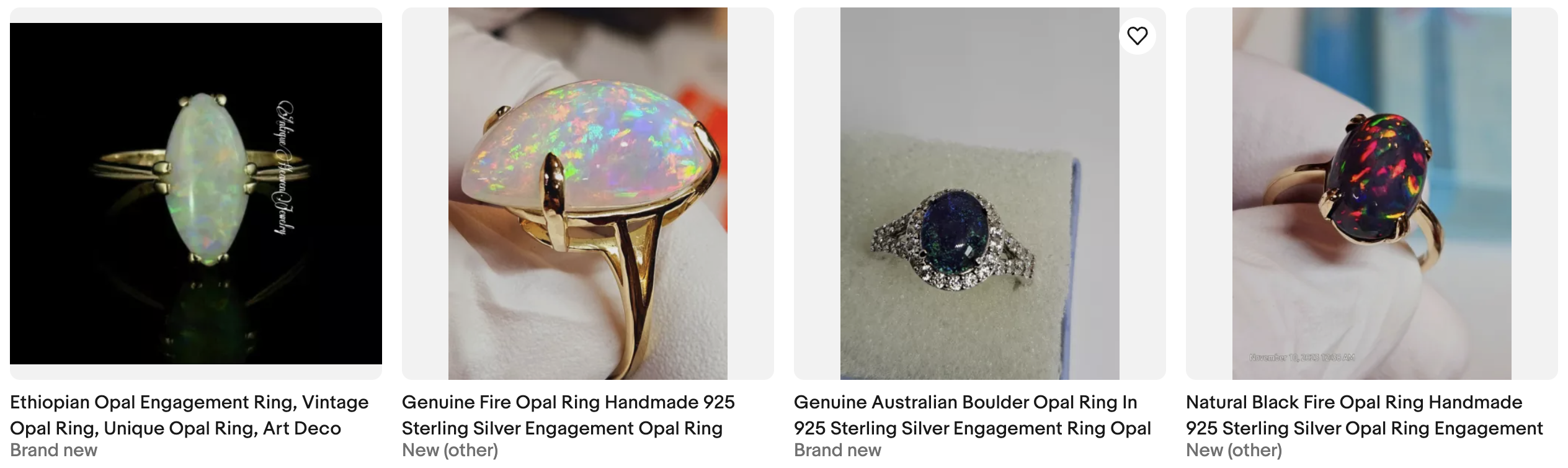}
        \caption{Products retrieved using the first title from NEAR$^2$@64.}
        \label{fig:errorsub2}
    \end{subfigure}
    \hfill
    \begin{subfigure}[b]{\columnwidth}
        \centering
        \includegraphics[width=\textwidth]{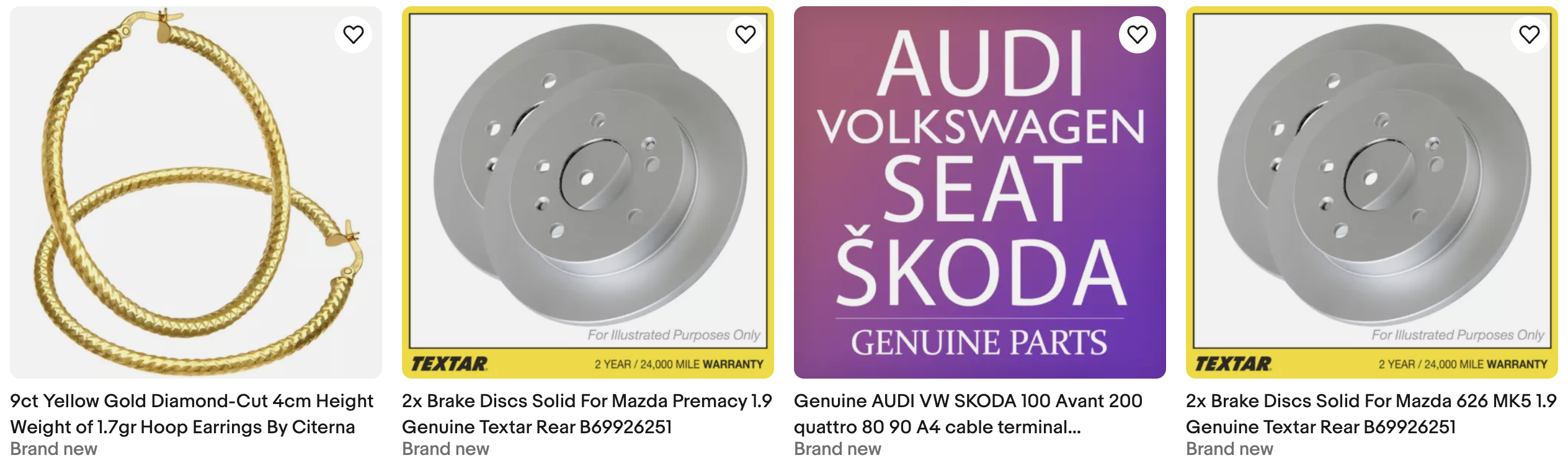}
        \caption{Products retrieved using the first title from eBERT without NEAR$^2$@64.}
        \label{fig:errorsub3}
    \end{subfigure}
    \caption{Products retrieved on eBay using the gold label title (a), the top one title from eBERT using NEAR$^2$@64 (b) and eBERT not using NEAR$^2$@64 (c) for the query-title pairs in Table~\ref{tab:error_example2}.}
    \label{fig:error_example2}
\end{figure}

\paragraph{Long and Detailed Query}
Table~\ref{tab:error_example2} presents the retrieved titles, their rankings, and their normalized similarity scores for the long and detailed query ``\textit{925 sterling silver triplet opal gemstone jewelry vintage pendant s-1.20}'' with eBERT. Given the specificity of the query, even using the exact gold label title did not yield the exact product on eBay. However, the model using NEAR$^2$@64 retrieved similar products, as shown in Figure~\ref{fig:error_example2}(b). In contrast, the products retrieved using top-ranked title from eBERT without NEAR$^2$@64, shown in Figure~\ref{fig:error_example2}(c), were significantly less relevant compared to those retrieved using the gold label title in Figure~\ref{fig:error_example2}(a). These results further demonstrate the effectiveness of NEAR$^2$@64. As with the short query example in Table~\ref{tab:error_example1}, normalized similarity scores from eBERT without using NEAR$^2$@64 are much lower than those using NEAR$^2$@64, further underscoring its limitations.

\begin{figure}[h]
  \centering
  \includegraphics[width=0.9\textwidth]{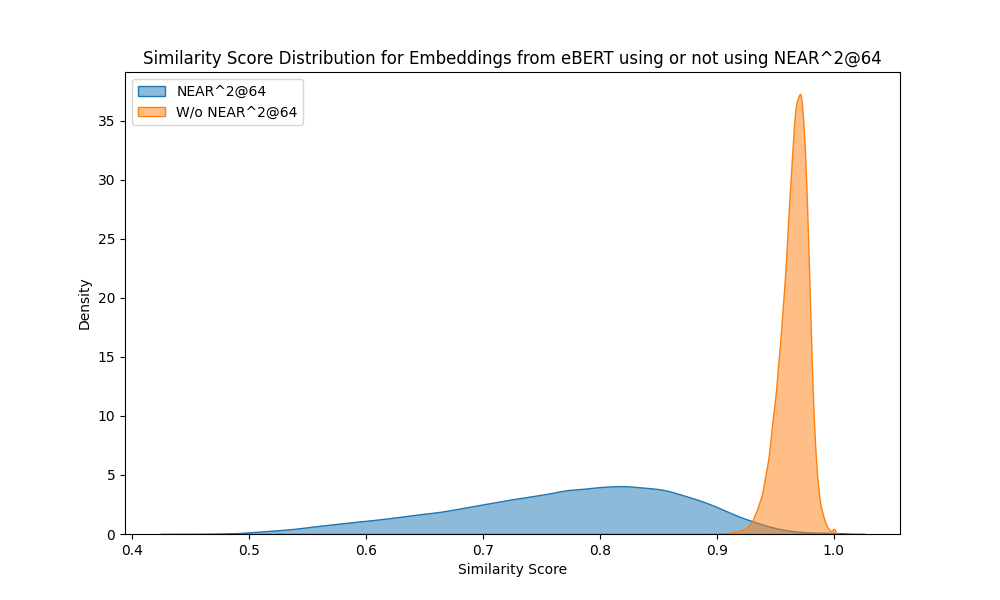}
  \caption{Similarity score distribution for embeddings from models \textbf{using} \textit{vs} \textbf{not using} NEAR$^2$@64  with eBERT on the CQ test set.}
  \label{fig.score_dis}
\end{figure}

\paragraph{Performance Disparity}
To investigate the root cause of performance disparity, we plotted the distribution of original similarity scores based on eBERT for all retrieved query-title pairs in the CQ test set, as shown in Figure~\ref{fig.score_dis}. The scores from the model using NEAR$^2$@64 are well-distributed between $0.5$ and $1.0$, reflecting nuanced relevance evaluations. In contrast, scores from eBERT without using NEAR$^2$@64 are clustered between $0.9$ and $1.0$, with most query-title pairs assigned a score near 0.95. This uniform distribution suggests that eBERT fails to effectively differentiate between relevant and irrelevant titles, leading to poor ranking performance. These findings further validate the superiority of NEAR$^2$@64 in the evaluation metrics for product retrieval and ranking tasks. 

\begin{figure}[h]
  \centering
  \includegraphics[width=0.9\textwidth]{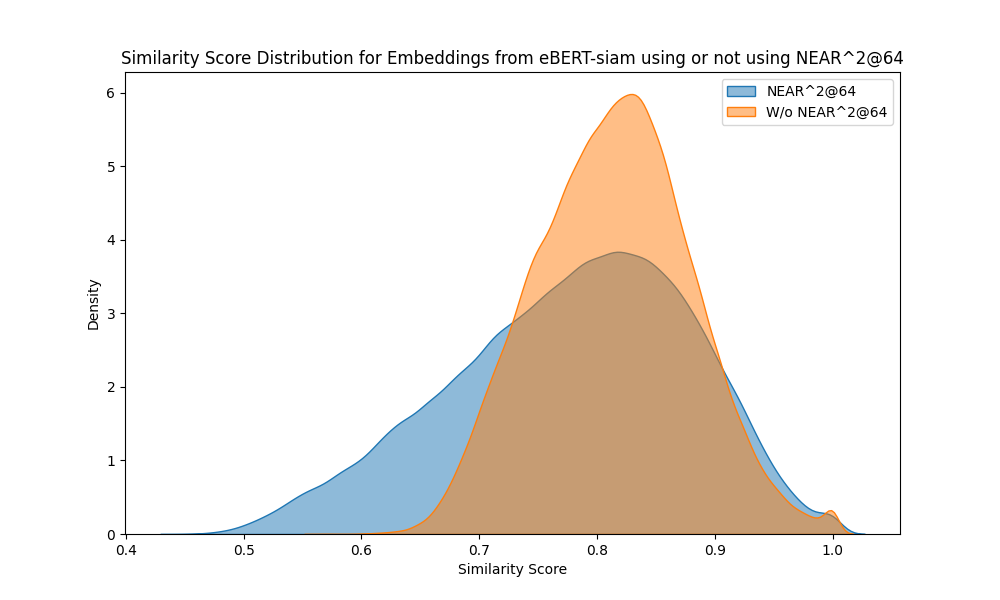}
  \caption{Similarity score distribution for embeddings from models \textbf{using} \textit{vs} \textbf{not using NEAR$^2$@64} with eBERT-siam on the CQ test set.}
  \label{fig.score_dis-siam}
\end{figure}

For product titles retrieved by eBERT-siam, whether for the short, implicit query or the long, detailed query, the differences in appearance between using and not using NEAR$^2$@64 are less pronounced compared to those observed with eBERT. However, the similarity scores still show a notable distinction. As illustrated in Figure~\ref{fig.score_dis-siam}, the model using NEAR$^2$@64 produces scores that are well-distributed between $0.45$ and $1.0$. In contrast, the scores from the model without this approach are more tightly clustered between $0.65$ and $1.0$, with the majority of query-title pairs receiving scores between $0.75$ and $0.9$. These results are consistent with the findings from the eBERT model. 

\begin{table*}[h]
\centering
\resizebox{15cm}{!}{%
\begin{tabular}{cccc}
\toprule
Method                         & Retrieved Title                                                                    & Ranking & Sim\_Score$_{Norm}$ \\ \midrule
\multirow{10}{*}{NEAR$^2$@64} &  CRAZY DAISY Shasta daisies  Qty 2  PLANTS  Hardy Perennial Healthy plants    & 1       & 0.3967           \\
                              &  CRAZY DAISY  Shasta daisies  Qty 2 x   Hardy Perennialhealthy plants      & 2       & 0.3824           \\
                              & Streptocarpus  MKsArktur09  young plant                                                  & 3       & 0.3822           \\
                              & Spathiphyllum Peace Lily Indoor Plants  1 x Potted Lily House Plant 9cm Pot       & 4       & 0.3731           \\
                              & Houseplant and Pot  Package                   & 5       & 0.3723           \\
                              & Spathiphyllum Peace Lily House Plant  Live Indoor House Potted Tree In 9cm         & 6       & 0.3710           \\
                              & Boston FernLive 10 Plants Lots Of Roots Air Purifier Reptile Terrarium ORGANIC      & 7       & 0.3696           \\
                              &  1 x CRAZY DAISY Shasta daisies  Hardy Perennial Healthy plant             & 8       & 0.3671           \\
                              & Leucanthemum Crazy Daisy Middleton Nurseries Flowering hardy Plants
                              & 9       & 0.3642           \\
                              & Syngonium White Butterfly Arrowhead Goose Foot Plant House Plant Easy Care        & 10      & 0.3640           \\ \hline
\multirow{10}{*}{W/o NEAR$^2$@64}  & Houseplant and Pot  Package                                                        & 1       & 0.2665           \\
                              & Spathiphyllum Peace Lily Indoor   Plants  1 x Potted Lily House Plant 9cm   Pot    & 2       & 0.2425           \\
                              & Spathiphyllum Peace Lily House Plant  Live Indoor House Potted Tree In 9cm         & 3       & 0.2417           \\
                              & Cordyline Kiwi  Ti Plant 7c Best Indoor Plants 7c Colourful   3040cm Potted Plant  & 4       & 0.2349           \\
                              & 68 Live Snake Plant Sansevieria   Trifasciata  Two Plants                          & 5       & 0.2341           \\
                              & Leucanthemum Crazy Daisy in plant in 13cm   pot approx                             & 6       & 0.2338           \\
                              & Multi Listing  Pond Plants Marginal Plants Water Bog   Garden Oxygenator SALE      & 7       & 0.2317           \\
                              & 12 Succulent Flowers not Included Pots 12   Pcs 12 Fashion Practical               & 8       & 0.2267           \\
                              & Avocado plant                                                                      & 9       & 0.2255           \\
                              & 3CM Succulent Cactus Live Plant Copiapoa   Tenuissima Chile Home Garden Rare Plant & 10      & 0.2239           \\ \hline
Gold label                    & Aloe Vera Plant - Large Plant in Pot                                               & /       & /      \\  \bottomrule         
\end{tabular}%
}
\caption{Retrieved titles for the detailed query ``plants'' using or not using NEAR$^2$@64 on \textbf{eBERT-siam}.}
\label{tab:error_example1siam}
\end{table*}

\begin{table*}[h]
\centering
\resizebox{15cm}{!}{%
\begin{tabular}{cccc}
\toprule
Method                         & Retrieved Title                                                                    & Ranking & Sim\_Score$_{Norm}$ \\ \midrule
\multirow{10}{*}{NEAR$^2$@64} & Moonstone Opal   Pendant 925 Sterling Silver Necklace Chain Womens Jewellery Gifts & 1       & 0.3934           \\
                              & Green Triplet Fire Opal Peridot 925 Sterling Silver Jewelry Pendants 27 v957 & 2       & 0.3814           \\
                              & Moonstone Opal Pendant 925 Sterling Silver Necklace Earring Women Jewellery Gift                & 3       & 0.3664           \\
                              & Sterling Silver 925 Signed Opal Heart Pendant Necklace 19 Chain                       & 4       & 0.3500           \\
                              & Vintage Possibly Opal Pendant On Gold Tone Necklace Chain         & 5       & 0.3337           \\
                              & Triplet Fire Opal Peridot Gemstone 925 Silver Jewelry Necklace 18 AQ269     & 6       & 0.3309           \\
                              & BULK LOT Vintage 925 Silver Costume Jewellery  Gemstones Opal Cloisonne Etc   & 7       & 0.3227           \\
                              & Ethiopian Opal 925 Sterling Silver Choker Necklace Women Gemstone Jewelry Gift                & 8       & 0.2829           \\
                              & Yellow Triplet Fire Opal Citrine 925 Sterling Silver Jewelry Earrings 21 s558          & 9       & 0.2691           \\
                              & Blue Opal Pendant 925 Sterling Silver  Minimalist Necklace Gift for Girlfriend & 10      & 0.2688           \\ \hline
\multirow{10}{*}{W/o NEAR$^2$@64}  & Vintage Possibly Opal Pendant On Gold   Tone Necklace Chain                        & 1       & 0.2615           \\
                              & Green Triplet Fire Opal Peridot 925   Sterling Silver Jewelry Pendants 27 v957     & 2       & 0.2561           \\
                              & Moonstone Opal Pendant 925 Sterling   Silver Necklace Chain Womens Jewellery Gifts & 3       & 0.2558           \\
                              & Triplet Fire Opal Peridot Gemstone 925   Silver Jewelry Necklace 18 AQ269          & 4       & 0.2505           \\
                              & Moonstone Opal Pendant 925 Sterling   Silver Necklace Earring Women Jewellery Gift & 5       & 0.2475           \\
                              & Sterling Silver 925 Signed Opal Heart   Pendant Necklace 19 Chain                  & 6       & 0.2472           \\
                              & GemporiaGems TV Sterling Silver 157ct   Ethiopian Blue Opal Pendant Necklace       & 7       & 0.2448           \\
                              & NWT GEMPORIA GEMS TV AUSTRALIAN OPAL   STERLING SILVER PENDANT                     & 8       & 0.2381           \\
                              & Vintage 925 Silver Opal Ring size J                                                & 9       & 0.2360           \\
                              & Australian Triplet Opal Gemstone 925   Sterling Silver Handmade Ring All Size      & 10      & 0.2270           \\ \hline
Gold label                    & 925 Sterling Silver Red Coral Gemstone   Handmade Jewelry Vintage Pendant S120     & /       & /      \\ \bottomrule         
\end{tabular}%
}
\caption{Retrieved titles for the detailed query ``925 sterling silver triplet opal gemstone jewelry vintage pendant s-1.20'' using or not using NEAR$^2$@64 on \textbf{eBERT-siam}.}
\label{tab:error_example2siam}
\end{table*}

\end{document}